\documentclass[aoas,MSNbibl,nameyear,seceqn,dvips]{arximspdf}
\usepackage{dcolumn}
\usepackage{graphicx}

\doi{10.1214/10-AOAS445}
\volume{5}
\issue{2B}
\pubyear{2011}
\firstpage{1611}
\lastpage{1631}

\makeatletter
\newcommand{\fraca}[2]{{#1/#2}}

\newcolumntype{d}[1]{D{.}{.}{#1}}
\newcommand{\eqref}[1]{(\ref{#1})}
\newtheorem{theorem}{Theorem}[section]

\renewcommand{\bm}[1]{\mathbf{#1}}
\newcommand{\btheta}{\bolds\theta}
\newcommand{\bV}{{\bm V}}
\newcommand{\by}{{\bm y}}

\def\bptnote#1{}

\makeatother

\begin{document}
\begin{frontmatter}

\title{Response-adaptive dose-finding under model uncertainty}
\runtitle{Response-adaptive dose-finding under model uncertainty}

\begin{aug}
\author[A]{\fnms{Bj\"{o}rn} \snm{Bornkamp}\ead[label=e1]{bornkamp@statistik.tu-dortmund.de}},
\author[B]{\fnms{Frank} \snm{Bretz}\ead[label=e2]{frank.bretz@novartis.com}},
\author[C]{\fnms{Holger} \snm{Dette}\corref{}\thanksref{a1}\ead[label=e3]{holger.dette@rub.de}}\\
\and~\author[D]{\fnms{Jos\'{e}}~\snm{Pinheiro}\ead[label=e4]{jpinhei1@its.jnj.com}}
\runauthor{Bornkamp, Bretz, Dette and Pinheiro}
\pdfauthor{Bjorn Bornkamp, Frank Bretz, Holger Dette, Jose Pinheiro}
\affiliation{TU Dortmund University, Novartis Pharma AG, Ruhr
University at~Bochum and Johnson \& Johnson}
\address[A]{B. Bornkamp\\
Technische Universit\"{a}t Dortmund\\
Fakult\"{a}t Statistik\\
44780 Dortmund\\ Germany\\\printead{e1}} 
\address[B]{F. Bretz\\
Novartis Pharma AG \\
Statistical Methodology \\
4002 Basel\\ Switzerland
\\\printead{e2}}
\address[C]{H. Dette\\
Ruhr-Universit\"{a}t Bochum \\
Fakult\"{a}t f\"{u}r Mathematik \\
44780 Bochum\\ Germany
\\\printead{e3}}
\address[D]{J. Pinheiro\\
Johnson \& Johnson \\
Raritan, New Jersey 08869\hspace*{9.2pt}\\
USA
\\\printead{e4}}
\end{aug}
\thankstext{a1}{Supported in part by the
Collaborative Research Center ``Statistical modeling of nonlinear
dynamic processes'' (SFB 823) of the German Research Foundation (DFG).}

\received{\smonth{4} \syear{2010}}
\revised{\smonth{11} \syear{2010}}

%
\begin{abstract}
Dose-finding studies are frequently conducted to evaluate the effect
of different doses or concentration levels of a compound on a~%
response of interest. Applications include the investigation of a
new medicinal drug, a herbicide or fertilizer, a molecular entity,
an environmental toxin, or an industrial chemical. In
pharmaceutical drug development, dose-finding studies are of
critical importance because of regulatory requirements that marketed
doses are safe and provide clinically relevant efficacy. Motivated
by a dose-finding study in moderate persistent asthma, we propose
response-adaptive designs addressing two major challenges in
dose-finding studies: uncertainty about the dose-response models and
large variability in parameter estimates. To allocate new cohorts
of patients in an ongoing study, we use optimal designs that are
robust under model uncertainty. In addition, we use a Bayesian
shrinkage approach to stabilize the parameter estimates over the
successive interim analyses used in the adaptations. This approach
allows us to calculate updated parameter estimates and model
probabilities that can then be used to calculate the optimal design
for subsequent cohorts. The resulting designs are hence robust with
respect to model misspecification and additionally can efficiently
adapt to the information accrued in an ongoing study. We focus on
adaptive designs for estimating the minimum effective dose, although
alternative optimality criteria or mixtures thereof could be used,
enabling the design to address multiple objectives. In an extensive
simulation study, we investigate the operating characteristics of
the proposed methods under a variety of scenarios discussed by the
clinical team to design the aforementioned clinical study.
\end{abstract}

%
\begin{keyword}
\kwd{Dose-response}
\kwd{drug development}
\kwd{minimum effective dose}
\kwd{optimal design}
\kwd{shrinkage approach}.
\end{keyword}

\end{frontmatter}

\section{Introduction}\label{sec:intro}

Dose-finding studies have several challenges in common. First, they
usually address two distinct objectives, which lead to different
requirements on the study design [\citet{rube1995a},
\citet{brethsupinh2008}]: (i) assessing evidence of a drug effect,
and (ii) estimating relevant target doses. Second, the form of the
dose-response relationship is unknown prior to the study, leading to
model uncertainty. This problem is often underestimated, although
ignoring model uncertainty can lead to highly undesirable effects
[\citet{chat1995}, \citet{drap1995}, \citet{hjor1994}]. Third, data
from dose-finding studies are usually highly variable. This issue is
of particular importance in pharmaceutical drug development, because
sample sizes are kept to a minimum for ethical and financial
reasons. It is therefore critical to develop efficient dose-finding
study designs that use the limited information as efficiently as
possible, while addressing the above challenges.

Many approaches have been proposed in the optimal design literature to
distribute patients efficiently with regard to given study objectives;
see \citet{wu1988}, \citet{fedoleon2001} and
\citet{kingwong2004}, among many others. However, most of this
work has concentrated on an assumed fixed dose-response model. As there
is typically considerable
model uncertainty at the planning stage of a dose-response study, these
methods have
limited practical use. Based on concepts introduced by
\citet{laeu1974} [see also \citet{cookwong1994},
\citeauthor{zhuwong2000}  (\citeyear{zhuwong2000,zhuwong2001}),
\citet{bieddettpepe2006}],
\citet{dettbretpepepinh2008} investigated model-robust
designs that provide efficient target dose estimates for a set of
candidate dose-response models, rather than for a single dose-response
model. However, their designs require knowledge about the unknown
parameters associated with the anticipated dose-response models as
well as the prespecification of model probabilities.

A natural remedy is to investigate response-adaptive designs
(adaptive designs, in short) with several
cohorts of subjects.
After each stage the accumulated
information of the ongoing study is used to update the
initial information of the underlying model parameters and model probabilities,
which in turn is used to calculate the design
for the subsequent stage(s). Several adaptive designs have
been developed for this problem; see, for example, \citet{millguildett2007}
and \citet{draghsuapadm2007} for recent
approaches using optimal design theory, or
\citet{zhoujosewolf2003},
\citet{muelberrgrie2006}, and \citet{waththal2008}
for recent Bayesian adaptive designs. \citet{drag2010}
performed an extensive simulation study that
compared five different adaptive dose-finding methods.

In this paper we propose adaptive designs addressing the three major
challenges described above: multiple study objectives, model
uncertainty and large variability in the data. For this purpose we
use the model-robust designs proposed by
\citet{dettbretpepepinh2008} together with a Bayesian
shrinkage approach to stabilize the parameter estimates, especially in the
early part of a study. This allows one to calculate parameter estimates
as well
as model probabilities that can then be used to calculate model-robust
designs for the subsequent stage(s) of the study. The resulting
designs are robust with respect to model misspecification and
additionally adapt to the continuously accrued information in an
ongoing study. We focus on adaptive designs for estimating the minimum
effective dose (MED), that is, the smallest dose achieving a clinically
relevant benefit over the placebo response. However, alternative
optimality criteria or mixtures of optimality criteria could be used,
enabling the design to address multiple objectives.

\section{Asthma dose-finding study}\label{sec:example}

The research for this article was motivated by a Phase II dose-finding
study for the development of a new pharmaceutical compound in
asthma. This was a multi-center, randomized, double-blind, placebo
controlled, parallel group study in patients with moderate persistent
asthma, who were randomized to one of seven active dose levels or
placebo. The primary endpoint was change from baseline in a lung
function parameter (forced expiratory volume in 1 second, $FEV_1$)
after 28 days of administration, scaled such that larger values
indicated a better outcome. The objective of the trial was to evaluate
the dose effects over placebo for the primary endpoint and to assess
whether there was any evidence of a~drug effect. Once such a
dose-response signal had been detected, one would subsequently
estimate relevant target doses, where the primary focus was on
estimating the MED.

Based on discussions with the clinical team, a homoscedastic
normal model was assumed for the primary endpoint with a standard
deviation of
350 ml, a placebo effect of 100 ml and a maximum treatment effect of
300 ml
within the dose range [0, 50] under investigation. The available doses
were 0 ($= {}$placebo), 0.5, 1.0, 2.5, 5, 10, 20 and 50. The
clinically relevant benefit over the placebo effect was set to
200 ml. That is, an increase in treatment effect of less than 200 ml over
the observed placebo response was considered to be clinically
irrelevant. Furthermore, all dose levels within the investigated dose
range were considered safe based on previous studies, so that efficacy
was of primary interest.

\begin{figure}[b]
\vspace*{-3pt}
\includegraphics{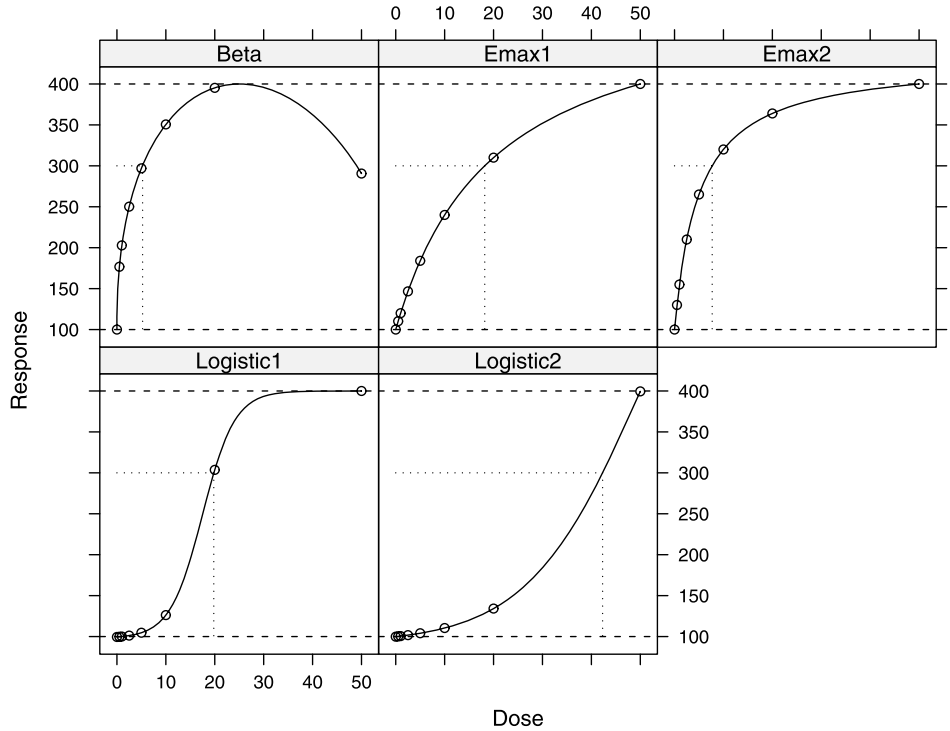}

\caption{Graphical display of the dose-response models in
Table~\protect\ref{tab1}. Open dots
denote the potential responses at the seven active dose levels and
placebo available in the
study. Dotted horizontal lines indicate the clinical relevance
threshold on top
of placebo response and dotted vertical lines the resulting MED.}
\label{fig1}
\end{figure}

Because this study was conducted early in the drug development
program, limited information about the dose-response shape was
available at the planning stage. A set of candidate dose-response
models was derived before starting the study; see Table~\ref{tab1} and
Figure~\ref{fig1} for the full model specifications (including a
preliminary specification of the model parameters). An increase of
the dose-response curve in the lower part of the investigated dose
range was considered likely, so two concave increasing models
(Emax$_1$, Emax$_2$) were included in the model set. In addition,
$S$-shaped (Logistic$_1$), unimodal (Beta) and convex (Logistic$_2$)
models were included in the candidate
model set to robustify the
statistical analysis with respect to model uncertainty. We refer to
\citet{pinhbornbret2006} for details on the use of candidate models
in dose-response studies and the elicitation of best guesses for the
model parameters.

\begin{table}[b]
\tabcolsep=0pt
\caption{Candidate dose response models as a function of dose $d$,
where $B({a},{b}) = ({a}+{b})^{{a}+{b}}/({a}^{a}{b}^{b})$}
\label{tab1}
\vspace*{-5pt}
\begin{tabular*}{\textwidth}{@{\extracolsep{\fill}}lccd{2.2}@{}}
\hline
\textbf{Model}&\textbf{Full model specification}&\textbf{Model parameters}&\multicolumn{1}{c@{}}{\textbf{True MED}}\\
\hline
Beta & $\theta_0+\theta_1B(\theta_2,\theta_3)(d/60)^{\theta_2}
(1-d/60)^{\theta_3}$ & (100, 300, 0.43, 0.6)&5.21\\
Emax$_1$ & $\theta_0 + \theta_1d / (\theta_2 + d)$ & (100, 420, 20)&
18.18\\
Emax$_2$ & $\theta_0 + \theta_1d / (\theta_2 + d)$ & (100, 330, 5)&
7.69\\
Logistic$_1$ & $\theta_0 + \theta_1/ \{1+\exp[(\theta_2 -
d)/\theta_3] \}$ & (98, 302, 17.5, 3.3)& 19.82\\
Logistic$_2$ & $\theta_0 + \theta_1/ \{1+\exp[(\theta_2 -
d)/\theta_3] \}$ & (92, 615, 50, 11.5)& 42.28\\
\hline
\vspace*{-8pt}
\end{tabular*}
\end{table}

Given the information and constraints above, the clinical team was
faced at the planning stage with several remaining key questions on the
study design:
\begin{longlist}[(B)]
\item[(A)] Should an adaptive design be employed at all or
would a nonadaptive design be sufficient?
\item[(B)] If the decision was to
employ an adaptive design, how many interim analyses should be
conducted?
\item[(C)] How many dose levels should be included in the study,
that is, are all seven active dose levels from above needed?
\item[(D)] If not
all active dose levels were needed, which of them should then be
investigated?
\end{longlist}
In addition to these statistical questions, many further
considerations were discussed by the clinical team: adaptive designs
require more logistical effort to set up the repeated data collection
and cleaning/analysis processes than nonadaptive designs; including
all seven active doses in the study would pose serious challenges to
the drug manufacturing and supply departments, especially if the
allocation changed during the study course; and how to ensure trial
integrity and validity. In the following we focus on the statistical
questions and describe the proposed methodology for the study.

\section{Methodology}
\label{sec:meth}
Assume $k$ distinct dose levels $d_1, \ldots , d_k$, where $d_1
= 0$ denotes placebo.
Let $n_i$ patients be allocated to dose $d_i$ and $N = \sum_{i=1}^k n_i$.
The vector of allocation weights is denoted by $\bm w=(w_1, \ldots ,
w_k)'$, where
$w_i=n_i/N$.
Let further $Y_{ij} \sim N(f(d_i, \btheta), \sigma^2)$
denote the observation of patient $j=1, \ldots , n_i$
at dose $d_i, i=1, \ldots , k$, where the dose-response model
$f(\cdot)$ is parameterized through the parameter vector
$\btheta$ and $N(\mu, \sigma^2)$ denotes the normal distribution
with mean $\mu$ and variance $\sigma^2$.

Most dose-response models
used in practice, including those in Table~\ref{tab1}, can be
decomposed as
%
\begin{equation}
\label{eqn:stand}
f(d, \btheta) = \theta_0 + \theta_1f^0(d, \bolds\theta^0),
\end{equation}
where $\btheta= (\theta_{0},\theta_{1},{\btheta^0}^{\prime
})^{\prime} =
(\theta_{0}, \ldots , \theta_{p})^{\prime}$. The parameters $\bolds
\theta^*=(\theta_{0},\theta_{1})'$ enter the model function $f$
linearly and determine its location and scale,
while~$f^0$ is typically a nonlinear function that
determines the shape of the model function $f$ through the
parameters $\btheta^0$.

The minimum effective dose producing a clinically relevant effect
$\Delta$
over the placebo response is defined as
%
\begin{equation}
\label{eqn:med}
\mathrm{MED}=\min_{d\in(d_1, d_k]}\{f(d,\bolds\theta
)>f(d_1,\bolds\theta)+\Delta\},
\end{equation}
where we assume that a beneficial effect is associated with larger values
of the response variable. Note that the MED may not exist, as no
dose in $(d_1, d_k]$ may produce an improvement of $\Delta$ compared with
placebo.

\subsection{Robust designs for MED estimation}
\label{ssec:optdes}


Given a function $f^0$, it follows from (\ref{eqn:med}) that the
MED (provided it exists) is a solution to
%
\begin{equation}
\theta_{0}+\theta_{1}f^0(0,\bolds\theta^0)+\Delta=\theta_{0}+\theta
_{1}f^0(\mathrm{MED},\bolds\theta^0).
\end{equation}
Consequently, $\mathrm{MED} = b(\bolds\theta) = h^0(f^0(\mathrm{0},\bolds
\theta^0)+\Delta/\theta_{1})$, where $h^0(x)=\inf\{ z|\break f^0(z)\geq x
\}$ denotes the (generalized) inverse of the function $f^0$ with
respect to the variable $d$. Standard asymptotic theory for nonlinear
models [\citet{sebewild1989}] yields that the maximum likelihood (ML)
estimate~$\widehat{\bolds\theta}$ is approximately multivariate normal
distributed with mean vector $\bolds\theta$ and covariance matrix
$\frac{\sigma^2}{N}\bm M^{-1}(\bolds\theta,\bm w ),$ where $\bm M(\bolds
\theta,\bm w )=\sum_{i=1}^kw_ig(d_i,\bolds\theta)g(d_i,\bolds\theta)'$
denotes the information matrix and $g(d,\bolds\theta)=\frac{\partial
f(d,\bolds\theta)}{\partial\bolds\theta}$ the gradient of the
dose-response model $f$ with respect to $\btheta$. It follows from the
$\delta$-method [see \citet{vand1998}] that the $\mathrm{MED}$ estimator
based on $\widehat{\bolds\theta}$,
$\widehat{\mathrm{MED}}=b(\widehat{\bolds\theta})$, is asymptotically
normally distributed with mean $b(\bolds\theta)$ and variance $V(\bolds
\theta, \bm w)=\frac{\sigma^2}{N}\bolds\nabla b(\bolds\theta)'\bm
M^{-1}(\bolds\theta,\bm w )\bolds\nabla b(\bolds\theta)$, where $\bolds\nabla
b(\bolds\theta) = \frac{\partial b(\bolds\theta)}{\partial\bolds\theta}$.
Hence, minimi\-zing $V(\bolds\theta, \bm w)$ with respect to $\bm w \in
\mathbb{S}^{k}=\{\bm w|\sum_{i=1}^kw_i=1, \bm w \geq0\}$ results in
optimal designs that minimize the approximate variance of
$\widehat{\mathrm{MED}}$. This design criterion has also an appealing
decision theoretic justification: The asymptotic normal distribution
of $\widehat{\mathrm{MED}}$ approximates the posterior distribution of
the $\mathrm{MED}$ in a Bayesian model framework. Hence, minimizing the
log-variance of $\widehat{\mathrm{MED}}$ is equivalent to minimizing
the (approximate) Shannon entropy of the posterior distribution of the
$\mathrm{MED}$ [\citet{chalverd1995}].

In principle, the above optimization could be done with respect to the
number and choice of doses and their corresponding allocation
ratios [\citet{dettbretpepepinh2008}], but, in practice,
manufacturing constraints often determine the available doses, as it
was the case in the asthma study from Section~\ref{sec:example}.
In the following we thus restrict the optimization to the weights $\bm
w$ for prespecified doses $d_1, \ldots , d_k$.


The true dose-response function $f$ is unknown and optimal designs are~%
ty\-pically not robust with respect to model misspecification
[\citet{dettbretpepepinh2008}]. In the following we assume a
set of $M$ candidate models $f_m(d,
\btheta_m)=\theta_{0m}+\theta_{1m}f_m^0(d,\bolds\theta^0_m)$, $m=1,
\ldots , M$, such as those described in Table~\ref{tab1}. We
``integrate'' the design criterion conditional on model $m$ with
respect to the model probabilities $\alpha_m$. Hence, using the design
criterion $\sum_{m=1}^M\alpha_m \log(V_m(\bolds\theta_m, \bm w))$ or,
equivalently,
%
\begin{equation}
\label{eqn:optCrit2}
\Psi(\bm w) = \prod^M_{m=1}(V_m(\bolds\theta_m, \bm w))^{\alpha_m}
\end{equation}
leads to designs that are robust with respect to model
misspecification, where $V_m(\bolds\theta_m, \bm w)$ denotes the
variance of the estimate for the MED in the $m$th model
$(m=1,\ldots ,M)$. Note that because of taking logarithms above, there is
no need to standardize the individual model variances. Otherwise this
would be necessary to avoid that some models dominate the design
criterion [the $V_m(\bolds\theta_m, \bm w)$ can be quite model dependent
and differing in size]. However, the numerical calculation of robust
designs using the criterion \eqref{eqn:optCrit2} requires the
knowledge of $\btheta_m$ and $\alpha_m, m=1, \ldots , M$. In the
following sections we describe how the initial best parameter guesses
can be updated during an ongoing study such that subsequent stages can
be redesigned based on the updated estimates for $\btheta_m$ and
$\alpha_m, m=1, \ldots , M$; see Section~\ref{ssec:algo} for a
description of the complete procedure in an algorithmic form.

\subsection{Updating of model parameters and weights}
\label{ssec:upd}

Reliably estimating the parameters $\bolds\theta_1, \ldots , \bolds
\theta_M$ is a challenging problem, particularly in early stages of a
study. ML estimates for these parameters are typically highly
variable, and may even not exist without imposing bounds on the
parameter space. One way of stabilizing estimates is to use a
shrinkage approach based on, for example, penalized maximum likelihood
or maximum a-posteriori (MAP) estimates. Here, one optimizes the
log-likelihood function plus a~term which determines the prior
plausibility of the parameters (the log prior distribution). The
estimate is then a compromise between the information contained in
data and the prior distribution. This stabilizes the estimates in
early stages due to the shrinkage toward a priori reasonable
values. In later stages the shrinkage effect decreases because the log
prior remains constant while the log likelihood receives more weight
with increasing sample sizes. If a~completely flat prior distribution
is used, standard ML and MAP estimation coincide, so that using
nonuniform priors is desirable. We discuss the choice of nonuniform
priors in more detail further below.



Apart from stable parameter estimates $\bolds\theta_1, \ldots , \bolds
\theta_M$ for the dose-response models, one needs to update the
model probabilities $\alpha_1, \ldots , \alpha_M$ at an interim analysis.
We propose using a probability distribution over the different
dose-response models and evaluating the posterior probabilities
for each model after having observed the data; see, for example,
\citet{kassraft1995} for a detailed description of
posterior probabilities and Bayes factors. These posterior model
probabilities can then be used in the design criterion
\eqref{eqn:optCrit2}. A computationally efficient approach to
approximate the posterior model probabilities is the Bayesian
information criterion (BIC). However, previous simulation studies
in the context of dose-finding studies showed that the BIC
approximation frequently favors too simplistic models for
realistic variances and sample sizes [see \citet{born2006}]. Other
approximate methods, such as fractional Bayes factors,
or intrinsic Bayes factors [see \citet{ohag1995} or \citet{bergperi1996}],
either depend on arbitrary tuning parameter values or are
computationally prohibitive. Thus, for each model we will use the exact
posterior
probabilities resulting from the prior distributions
assumed for the MAP estimation.
In our case, these probabilities can be calculated using
efficient numerical quadrature without the
need to resort to computationally expensive Markov chain Monte
Carlo techniques. In the remainder of this section we provide
details on the prior elicitation and the calculation of posterior
probabilities.

\subsubsection{Selection of prior distributions for $\bolds\theta_m$}
\label{sssec:priors}

We utilize the factorization in (\ref{eqn:stand}) to derive a
prior distribution for $(\theta_{0m},\theta_{1m},\bolds\theta^0_m,
\sigma^2)$. If $\bolds\theta^0_m$ were known, the nonlinear
models would reduce to a linear model. It is therefore reasonable
to use for a given $\bolds\theta^0_m$ the conditionally conjugate
normal-inverse gamma (NIG) distribution
\[
p(\btheta^*_m,\sigma^2|\bolds\theta^0_m)\propto(\sigma^2)^{-(\nu
+4)/2}\exp[-\{(\btheta^*_m-\bolds\mu)'\bV^{-1}(\btheta^*_m-\bolds\mu
)+a\}/(2\sigma^2)]
\]
for $(\btheta^*_m,\sigma^2)$ [see \citet{ohagfors2004}], where the
parameter $\btheta^*$ is defined after equation \eqref{eqn:stand},
$a,\nu>0$, $\bolds\mu\in\mathbb{R}^2$ and $\bm V \in
\mathbb{R}^{2\times2}$ denotes a positive definite matrix. The NIG
distribution marginally induces a bivariate $t$-distribution for
$\btheta^*_m$ with $\nu$ degrees of freedom, finite mean $\bolds\mu$ and
covariance matrix $a/(\nu-2)\bm V$, provided that $\nu>2$. The
marginal prior distribution for~$\sigma^2$ is given by an IG
distribution with mode $a/(\nu+2)$, mean $a/(\nu-2)$ and variance
$2a^2/\{(\nu-2)^2(\nu-4)\}$. It has a~finite mean when $\nu>2$ and a~%
finite variance when $\nu>4$.

To set up the NIG distribution for $(\btheta^*_m, \sigma^2)$, one can
employ available information about the placebo effect, the maximum
treatment effect and the standard deviation. For example, one can
choose the marginal bivariate $t$-distribution for $\btheta^*_m$
(conditional on $\btheta^0_m$) such that the desired mean and
covariance are achieved for the placebo effect and the maximum effect
of the underlying dose-response model. When the linear parameters $\bolds
\theta^*_m$ cannot be interpreted as placebo and maximum
effect, one can use a suitable transformation to achieve the desired moments.
Then, one can adjust $a$ and~$d$ so that the marginal distribution of
$\sigma^2$ achieves the desired mode. An attractive choice is to use
$\nu=4$, leading to a prior with infinite variance for~$\sigma^2$ and a
heavy tailed marginal prior for $\btheta^*_m$.



\begin{figure}

\includegraphics{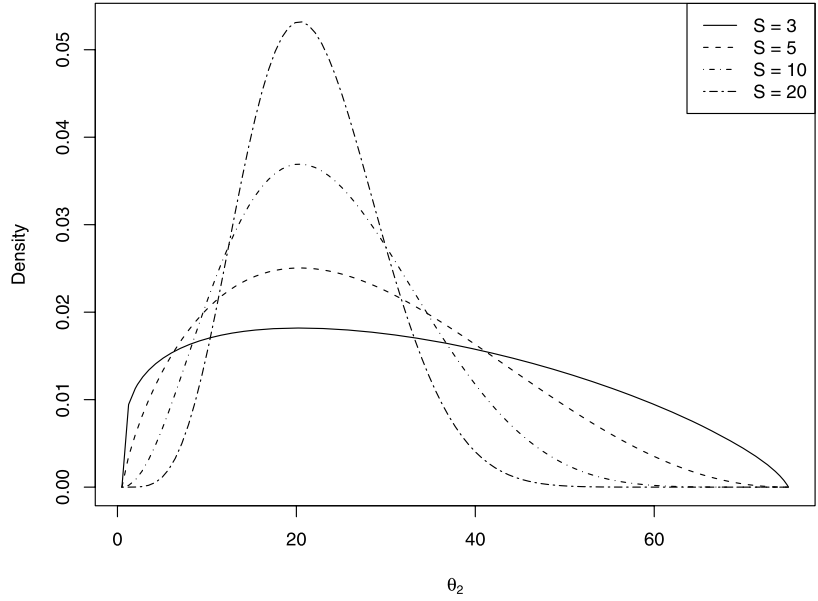}

\caption{Beta priors on $[0.5,75]$ with mode 20 and different $S$
values.}
\label{fig:beta}
\end{figure}

For the nonlinear parameters $\bolds\theta^0_m$, we propose
selecting suitable bounds for the parameters and then eliciting a
bounded prior distribution. This is typically not difficult,
as the interpretation of the nonlinear parameters is straightforward, and
excessively large parameter values usually correspond to a priori
unlikely model shapes. We propose using a scaled beta distribution
$B(\alpha,\beta)$ with mode equal to the initial parameter
guesses. The curvature of the prior determines the amount of
shrinkage that one is willing to employ for the MAP estimates. In
the simulations we used the sum $S=\alpha+\beta$
as a~measure of curvature with $S>2$ to ensure
unimodality of the beta distribution.
Note that already relatively small values of~$S$, such as $S= 10$
or $S=20$, lead to strong shrinkage effects; see
Figure~\ref{fig:beta} for an illustration of the $\theta_2$
parameter in the Emax$_1$ model, where the initial parameter guess
is~20. For dose-response models with more than one nonlinear
parameter, we repeat this procedure for all parameters and assume
independence among them.\looseness=-1


For selecting prior model probabilities,
it is convenient to use a uniform distribution across the models
unless some models are deemed a priori more plausible than others.



\subsubsection{Calculation of posterior probabilities}
\label{sssec:implem}

Let $\bm y$ denote the data available at an interim analysis and
$p(\by|\btheta_m,m)$ the likelihood under model $m$ with corresponding
prior distribution $p(\btheta_m|m)$ and prior model probabili\-ty~%
$p(m)$. Then the marginal likelihood is given by
\begin{eqnarray*}
p(m|\by)&\propto&
p(m)\int p(\by|\btheta_m,\sigma^2,m)p(\btheta_m, \sigma^2|m)\,
d(\btheta_m, \sigma^2) \\
&\propto&
p(m)\int\int p(\by|\btheta_m,\sigma^2,m)p(\btheta^*_m,\sigma
^2|\bolds\theta^0_m,m)\, d(\btheta^*_m,
\sigma^2)p(\bolds\theta^0_m|m)\,
d\bolds\theta^0_m.
\end{eqnarray*}
The inner integral in the last equation is the product of a likelihood
and a~con\-jugate prior distribution. One can hence reduce the
integration to
%
\begin{equation}
\label{eqn:mlh3}
p(m|\by)\propto p(m)\int
p(\by|\btheta^0_m,m)p(\bolds\theta^0_m|m)\, d\bolds\theta^0_m,
\end{equation}
where $p(\by|\btheta^0_m,m)$ now denotes the integrated likelihood. In
our applications, the integral (\ref{eqn:mlh3}) is one- or
two-dimensional over a bounded region and hence straightforward to
calculate numerically. This allows us to calculate the marginal
likelihoods efficiently, without resorting to Markov chain Monte Carlo
calculations; see Section~\ref{ssec:details} for details. The
posterior model probabilities $p(m|\by)$ can be obtained by
normalizing the marginal likelihoods (multiplied by the prior model
probabilities).

We use the maximum $\tilde{\bolds\theta}^0_m$ of the marginal posterior
$p(\by|\btheta^0_m,m)p(\bolds\theta^0_m|m)$ as an estimate of $\bolds
\theta^0_m $. Conditional on this value, we use the maximum
$\tilde{\btheta}^*_m$ of $p(\btheta^*_m|\tilde{\btheta}^0_m, \bm y,m)$
as an estimate for $\btheta^*_m$. Therefore, the overall estimate of
the parameter $\bolds\theta_m$ is given by
$\tilde{\btheta}_m=(\tilde{\btheta}^*_m, \tilde{\btheta}^0_m)$. This
is a slight variation of the MAP approach described above, but reduces
further the computational effort, as it reuses the calculations from
the integration to obtain the marginal likelihoods.

\subsection{Main algorithm}
\label{ssec:algo}

We now summarize the complete response-adaptive dose-finding
design in algorithmic form.

\textit{Before trial start}:
\begin{longlist}[(5)]
\item[(1)] Select a starting design using either a balanced allocation
across the available doses or an unbalanced allocation based on optimal
design considerations.
\item[(2)] Select candidate dose-response models $f_m(d,\bolds\theta_m)$.
\item[(3)] Conditional on $\bolds\theta_m^0$, calculate a NIG prior
distribution for ($\bolds\theta^*_m$, $\sigma^2$) based on ``best
guesses'' for the placebo effect, the maximum treatment effect and
$\sigma^2$ (together with suitable variability assumptions for
both parameters).
\item[(4)] Choose ``best guesses'' for the nonlinear parameters
$\btheta^0_m$ and select the parameter $S$.
\item[(5)] Choose prior model probabilities $p(m)$ for the different
dose-response functions.
\end{longlist}

\textit{At interim analysis}:
\begin{longlist}[(3)]
\item[(1)] Calculate posterior model probabilities
%
\begin{equation}
\label{eqn:modprob}
p(m|\by)\propto p(m)\int
p(\by|\btheta_m,m)p(\btheta_m|m)\,d\btheta_m.
\end{equation}
Exploiting the conjugacy properties of the NIG distribution,
this reduces to one- or two-dimensional
integrals; see Section~\ref{ssec:details} for computational
details.
\item[(2)] For each dose-response
model, estimate ${\btheta}_m$ by using the maximum of $p(\by|\btheta
^0_m,m)p(\bolds
\theta^0_m|m)$, where the abscissas calculated in step 1 can be reused.
Conditional on this value, use
the maximum of $p(\btheta^*_m|\tilde{\btheta}^0_m, \bm y,m)$ as an
estimate for $\bolds\theta^*_m$ to obtain $\tilde{\btheta}_m$
\item[(3)] Plug the obtained parameter estimates $\tilde{\btheta}_m$ into
\eqref{eqn:optCrit2}
and set $\alpha_m = p(m|\by)$. Then, minimize with respect to
$\bm w \in\mathcal{S}$, where $\mathcal{S}=\{\bm w
\in\mathbb{S}^k | \bm w = (\bm n_{\mathrm{old}}+ N_{\mathrm{next}}\bm
w_{\mathrm{next}})(N_{\mathrm{old}}+N_{\mathrm{next}})^{-1}, \bm w_{\mathrm{next}} \in
\mathbb{S}^k\}$. Here, $\bm n_{\mathrm{old}}$ denotes the vector of sample
sizes per dose and $N_{\mathrm{old}}$ the total sample size until the
current interim analysis. Further, $N_{\mathrm{next}}$ denotes the sample
size and ${\bm w}_{\mathrm{next}} \in\mathbb{S}^k$ the design weights for
the next cohort of patients. We therefore optimize the design for
the next stage taking into account the patient allocation until
the current interim analysis; see Section~\ref{ssec:details} for
computational details.
\item[(4)] Allocate the next cohort of patients according to $\bm
w_{\mathrm{next}}$ by applying an appropriate rounding technique, such as
described in \citet{puke1993}, Chapter~12.
\end{longlist}

Note that the Bayesian approach is used here for design adaptation
purposes. The final analysis may or may not be done using a fully
Bayesian approach. The development of the Bayesian design
methodology above is motivated by the MCP-Mod methodology described in
\citet{bretpinhbran2005} to address model uncertainty. This method
requires prior estimates for the placebo effect,
the maximum treatment effect, and $\sigma^2$ at the design stage and ``best
guesses'' of the nonlinear parameters $\bolds\theta^0_m$ for the
analysis. The
additional information needed to set up the above adaptive design
procedure is hence minimal. Obviously, any other strategy that allows
one to use a set of candidate dose-response models might also be used
for the final analysis.

\subsection{Technical details}
\label{ssec:details}

In this section we provide further details of the algorithm presented above.

For the calculation of the one- and two-dimensional integrals in
\eqref{eqn:mlh3} we used quasi-uniformly distributed point sets based
on good lattice points $\bm u_1, \ldots , \bm u_n$, where $\bm u_i \in
[0,1]^d$ and $d$ is the dimension of $\bolds\theta_m^0$; see
\citet{fangwang1994} for details on the construction of such
integration grids. Let $\pi(\bolds\theta_m^0|y)=p(\bm y|
\bolds\theta_m^0,m)\times p(\bolds\theta^0_m|m)$ denote the integrand from
\eqref{eqn:mlh3} and let $\bm{b}_l$ and $\bm{b}_u$ denote the vector
of lower and upper bounds for $\bolds\theta_m^0$. One first transforms
the good lattice points to obtain $\bm u_i^*=\bm u_i(\bm b_u-\bm
b_l)+\bm b_l$ for $i=1, \ldots , n$, and then approximates the integral
\eqref{eqn:mlh3} by $\prod_{j=1}^d(b_{uj}-b_{lj})\sum_{i=1}^n \pi
(\bm
u^*_i|y)/n$. This approach also allows one to calculate the
approximate maximum in the subsequent optimization step by using the
grid point $\bm u^*_i$ corresponding to $\max_{i} \pi(\bm
u^*_i|y)$. We found that using a grid of size 100 in the
one-dimensional case and the good lattice point set of size 1597 in
the two-dimensional case [\citet{fangwang1994}] provide sufficiently
reliable and computationally efficient results (both for integration
and optimization).

The optimization in \eqref{eqn:optCrit2} is a constrained optimization
problem because the weights $\bm w_{\mathrm{next}}$ lie in the
($k-1$)-dimensional probability simplex. A simple but efficient
approach to perform the optimization is to use a mapping
$\mathbb{R}^{k-1} \mapsto\mathbb{S}^k$ and then to employ a standard
unconstrained optimizer, as described in
\citet{atkidonetobi2007}, page~131. To account for the already
allocated patients until an interim analysis, one can optimize
$\Psi (\frac{\bm n_{\mathrm{old}}+N_{\mathrm{next}}\bm
w_{\mathrm{next}}}{N_{\mathrm{old}}+N_{\mathrm{next}}} )$ with respect to $\bm w_{\mathrm{next}}
\in\mathbb{S}^k$. Due to potential multiple optima in the design
surface, one cannot be sure whether indeed an optimal design has
been found by the optimizer. We thus propose using lower bounds of the
resulting relative efficiencies based on the underlying geometry of
the optimization problem.

To be precise, suppose that the vector
${\bm w}^*$ has been found by the optimizer. The following result
gives a lower bound on $r({\bm w}^*)=\frac{\Psi(\bm
w_{\mathrm{opt}})}{\Psi({\bm w}^*)}\in[0,1]$, where~$\bm w_{\mathrm{opt}}$ is the
(unknown) true optimal design at the end of the next stage,
accounting for the patients allocated until the current interim
analysis. A~proof of the result is given in the
\hyperref[appm]{Appendix}.



\begin{theorem}\label{theo1}
A design\vspace*{1pt} $\bm w$ with $ c_m(\bolds\theta_m ) =
\bolds\nabla b_m(\bolds\theta_m )\in\operatorname{Range} (\bm
M_m(\bolds\theta_m,\break
\bm w ))$, $m=1, \ldots , M$, minimizes $\Psi(\bm w)$ with respect to
$\bm
w_{\mathrm{next}}$, where $\bm w=\break\frac{ N_{\mathrm{old}}+N_{\mathrm{next}}\bm
w_{\mathrm{next}}}{N_{\mathrm{old}}+N_{\mathrm{next}}}$, if and only if there exist
generalized inverses $\bm G_1, \ldots , \bm G_m$ of $\bm M_m(\bolds
\theta_m, \bm w)$, such that the inequality
\[
\hspace*{-3pt}h(d,\bm w)\!=\!\frac{\sum_{m=1}^M\alpha_m\fraca{(g_m^T(d, \bolds
\theta_m)\bm G_mc_m(\bolds\theta_m))^2} {c_m^T(\bolds\theta_m)\bm
G_mc_m(\bolds\theta_m)}}{\sum_{m=1}^M\alpha_m\fraca{c_m(\bolds
\theta_m)\bm G_m^T\bm M_m(\bolds\theta_m,\bm w_{\mathrm{next}})\bm G_mc_m(\bolds
\theta_m)}{c_m^T(\bolds\theta)\bm G_mc_m(\bolds\theta)}} \!\leq\! 1
\]
is
satisfied for all $d \in\{d_1, \ldots , d_k\}$. Moreover, the
efficiency of any design
$\bm w$ can be bounded from below by
%
\begin{equation} \label{low}
r({\bm w}) \geq\frac{1}{k^*(\bm w, \gamma)} \ge \frac{1}{h^*(\bm w)},
\end{equation}
where $\gamma= \frac{N_{\mathrm{old}}}{N_{\mathrm{old}}+N_{\mathrm{next}}}$,
$ h^*(\bm w) =\min _{{\bm G_1, \ldots , \bm G_m}}\max_{d \in
\{d_1, \ldots , d_k\}} h(d,\bm w)$,
%
\begin{eqnarray*}
k^*(\bm w, \gamma) &=&
 \Biggl[ 1+(1-\gamma) \min_{\bm G_1, \ldots , \bm G_m} \max_{\bm v
\in\mathbb{S}^k} \sum^M_{m=1} \alpha_m
c^T_m(\bolds\theta_m)\bm G^T_m \\
&&\hphantom{
 \biggl[ 1+(1-\gamma) \min_{\bm G_1, \ldots , \bm G_m} \max_{\bm v
\in\mathbb{S}^k} \sum^M_{m=1}}\hspace*{-10pt}
{}\times\bigl(\bm M_m(\bolds\theta_m,\bm v)-\bm
M_m (\bolds\theta_m, \bm w_{\mathrm{next}})\bigr)\\
&&\hphantom{
 \biggl[ 1+(1-\gamma) \min_{\bm G_1, \ldots , \bm G_m} \max_{\bm v
\in\mathbb{S}^k} \sum^M_{m=1}}\hspace*{-10pt}
{}\times
\bm G_m c_m(\bolds\theta)
/(c^T_m (\bolds
\theta_m) \bm G_m
c_m (\bolds\theta_m) ) \Biggr]^{-1}\!,
\end{eqnarray*}
and the minimum is taken over
all generalized inverses $\bm G_1, \ldots , \bm G_m$ of the matrices
$\bm M_1(\bolds
\theta_1, \bm w), \ldots , \bm M_M(\bolds\theta_M, \bm w)$.
\end{theorem}

When the matrices $\bm M_1(\bolds\theta_1, \bm w), \ldots , \bm
M_M(\bolds\theta_M, \bm w)$ are invertible, $h^*(\bm w)$ is just the
maximum of $h(d,\bm w)$ over the $k$ doses and straightforward to
calculate [and so is the lower bound on $r(\bm w)$]. This lower
bound is useful in several respects: we do not need to know the actual
optimal design $\bm w_{\mathrm{opt}}$ in order to calculate the lower bound. If
the lower bound for our calculated design $\bm w^*$ is equal to
1, we know that $\bm w^*$ is the optimal design. Otherwise,
we have a~conservative estimate on
how much percent off one would be when using~$\bm w$. The bound
based on $k^*(\bm w, \gamma)$ is sharper, however harder to
implement.\looseness=-1

If one does not use a fully Bayesian approach for the final
analysis, one typically has to fit nonlinear regression models to
the data. When there are only few doses available,
as it is often the case in drug development practice, calculating
the ML estimate may be difficult. One way to simplify the problem
is by exploiting the fact that $\theta_0$ and $\theta_1$ enter the model
function linearly in \eqref{eqn:stand}. We thus apply the
nonlinear optimization only on the nonlinear parameters $\bolds
\theta^0_m$, similar in spirit to \citet{golupere2003}.
Using the Frisch--Waugh--Lovell theorem [\citet{balt2008}, Chapter~7], we
can recalculate the residual sum of squares efficiently, without
the need to solve the full least squares problems in each
iteration of the nonlinear optimization (this effect becomes even
more important when there are additional linear covariates in the
model equation, such as gender, baseline values, etc.). In
addition, we impose bounds on the nonlinear parameters $\bolds
\theta^0_m$ to guarantee the existence of the least squares
estimate [\citet{sebewild1989}, Chapter~12]. As mentioned in
Section~\ref{sssec:priors}, such bounds are not a severe
restriction in practice and ensure that the optimization problem
is well posed.

\section{Asthma study revisited}

In this section we revisit the asthma case study from
Section~\ref{sec:example} and address the four open design questions
using the proposed methodology from Section~\ref{sec:meth}.
To this end, we investigated in an extensive simulation study the operating
characteristics for different design options and parameter configurations.

\subsection{Design of simulation study}

We generated normally distributed observations according to the
dose-response models given in Table~\ref{tab1} with $\sigma=350$
ml. To investigate the robustness of the proposed methods, we also
simulated from a linear model (with baseline 100 ml and maximum effect
300 ml) that was not included in the candidate model set. The total
sample size was fixed at 300 (constraint imposed by the clinical
team). To evaluate the benefit of including additional doses, we
compared two design options, one with the four active doses 2.5, 10,
20, 50 (plus placebo) and another one with the seven active doses 0.5,
1, 2.5, 5, 10, 20 and 50 (plus placebo). In addition, we evaluated the
benefit of additional interim analyses by varying their number from 0
($= {}$no interim analysis) to 9, where the interim looks were chosen
equally spaced in time. In all cases, we assumed a balanced first
stage design. The designs from the second stage onward were
determined using the observed data according to the algorithm from
Section~\ref{ssec:algo}. When the MED estimate did not exist for
certain models at an interim analysis, they were removed from the
model set for the purpose of design calculation and the model
probabilities were reweighted accordingly. When the MED estimate did
not exist for any model, a balanced allocation was used for the next
cohort of patients.

For the final analysis we employed the MCP-Mod procedure from
\citet{bretpinhbran2005}. A potential dose-response signal was
assessed using model-based multiple contrast tests based on the
candidate model set from Table~\ref{tab1}. Subsequently, if there
were significant models, the dose-response model with lowest Akaike
Information Criterion (AIC) among the significant models was chosen to
estimate the MED.

The methodology from Section~\ref{sec:meth} was applied with uniform
prior probabilities for the different models. We further assumed
a priori distributions with mean 100 and variance 100,000 for the
placebo effect and mean 300 and variance 100,000 for the maximum
treatment effect, which were then transformed into the linear
parameters for all dose-response models. The mode of the marginal
distribution for $\sigma^2$ was chosen as $350^2$ with $\nu=4$,
resulting in an infinite variance. For the nonlinear parameters we
assumed beta distributions (or products thereof) with mode equal to
the values specified in Table \ref{tab1} and $S=3$. The parameter
bounds were chosen to ensure that all reasonable dose-response shapes
remained included within the bounds. That is, we chose $\theta_2 \in
[0.05,75]$ for the Emax models, $(\theta_2,\theta_3) \in
[0.5,4]\times[0.5,4]$ for the beta models and $(\theta_2,\theta_3)
\in
[0.05,75]\times[0.5,25]$ for the the logistic model. For each
scenario we used 5000 simulation runs.

\subsection{Simulation results}

For the chosen standard deviation of $\sigma= 350$~ml, the power of the
MCP-Mod procedure to detect a dose-response signal was almost always
close to 1. Thus, the MCP-Mod procedure was essentially reduced to
choosing the nonlinear model with lowest AIC value under the
constraint that only models with significant contrast test statistics
were included in the model selection step. Simulations with $\sigma$
values larger than $350$ ml indicated that the power quickly dropped to
lower levels (results not shown here), although the estimation results
remained qualitatively similar to the ones shown below.

\begin{figure}

\includegraphics{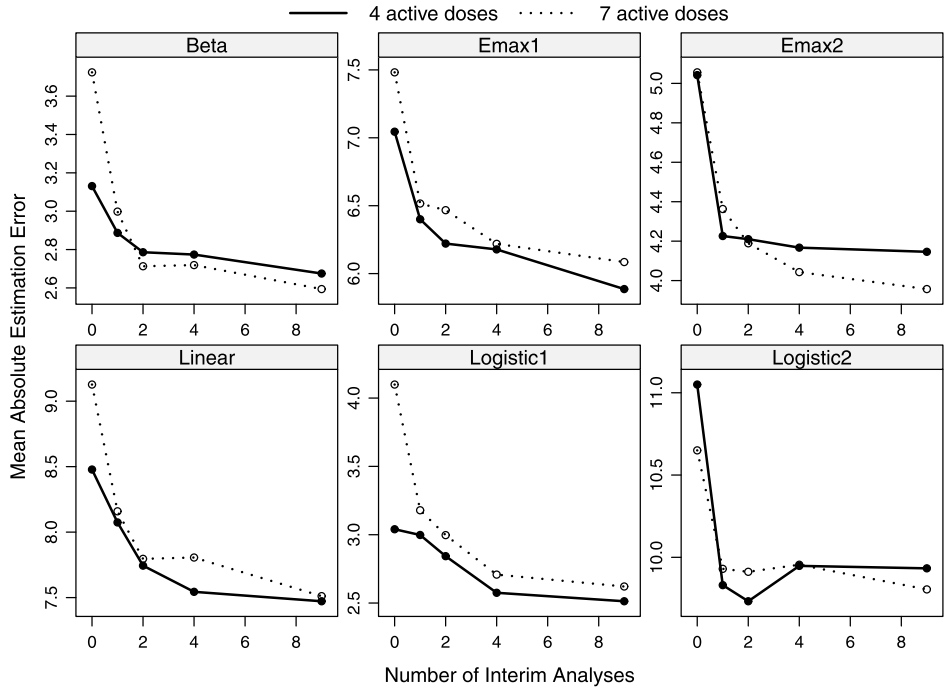}

\caption{Mean absolute estimation error for MED estimation.}
\label{fig2}
\vspace*{-5pt}
\end{figure}

In Figure~\ref{fig2} we display, for each simulation scenario, the
mean absolute estimation error for the MED against the number of
interim analyses. In all scenarios one observes a benefit from
adapting, while most of the improvement is already achieved after 1, 2
or 4 interim analyses. The largest relative improvement (comparing no
adaptations vs 9 adaptations) can be observed for the Logistic$_1$ and
the Beta model scenarios, particularly in the case of 7 active
doses. The worst relative improvement can be observed for the
Logistic$_2$ scenario, where the overall largest absolute estimation
error can be observed. This is not surprising, because even when
adapting one cannot achieve a good design for this model, as there are
no doses available for administration in the interval $(20,50)$
containing the MED; see also Figure \ref{fig1}. It is remarkable to
see that adaptation also works in the linear model scenario, although
the linear model is not included in the candidate model set. It seems
that other models in the candidate set are able to capture the shape
of a linear model reasonably well.

The comparison between 4 and 7 active doses is not entirely clear.
If no interim analyses are performed, it seems that the design with a
balanced allocation
across the 4 active doses is slightly better than the design with a~balanced
allocation across all 7 active doses. If one decides to adapt, however,
it seems beneficial in
some cases to have more doses available,
particularly if many interim analyses are performed, while in
other cases 4 active doses are sufficient.

\begin{figure}

\includegraphics{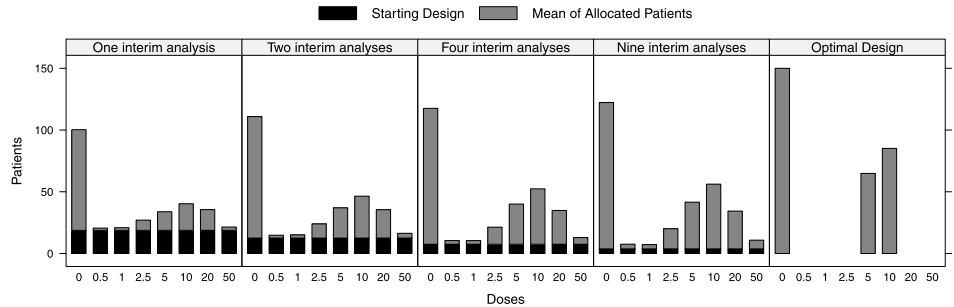}%
\vspace*{-5pt}
\caption{Average patient allocation after 1, 2, 4 and
9 interim analysis under the Emax$_2$ model. \textup{Last
panel:} locally MED-optimal design for the Emax$_2$ model with true MED${} = {}$7.7.}
\label{fig3}
\vspace*{-3pt}
\end{figure}

To illustrate how adaptation changes the allocation of patients to the different
doses, we display in Figure~\ref{fig3} the average patient allocations
for the Emax$_2$ model
after 1, 2, 4 and 9 interim analyses and with 7 available active doses.
The adaptive design tends to allocate more patients both on placebo
and nearby the actual MED. This is intuitively
plausible, as the MED estimate depends on the precision of the
estimated placebo effect as well as of the estimated function $f(\cdot)$
around the true MED.
It also follows from Figure~\ref{fig3} that for a large number of
interim analyses the overall allocation is close to the one under a
locally optimal design for the Emax$_2$ model, with the variability in
the allocations due to the uncertainty both in estimating the correct
model and the model parameters at the interim analysis. Similar
conclusions also hold for other models than the Emax$_2$ model (not
reported here).

We now investigate to which extent the precision gain observed in
Figure~\ref{fig2} translates into sample size savings when
performing an adaptive design. In other words, how many additional
patients are required for a nonadaptive, balanced design to
achieve a similar estimation error as with an adaptive
design using 300 patients. We again considered the Emax$_2$ model
and iterated the total sample size until the mean absolute
estimation error was approximately 4 (which is the mean absolute
estimation error obtained after 9 interim analyses, as seen in
Figure \ref{fig2}). For both design options with 4 and 7 active
doses, this was achieved after roughly 500 patients. Thus, using a
nonadaptive, balanced design, one would need 200 additional
patients to achieve a similar precision in MED estimation as
compared to an adaptive design using 300 patients.

\begin{figure}

\includegraphics{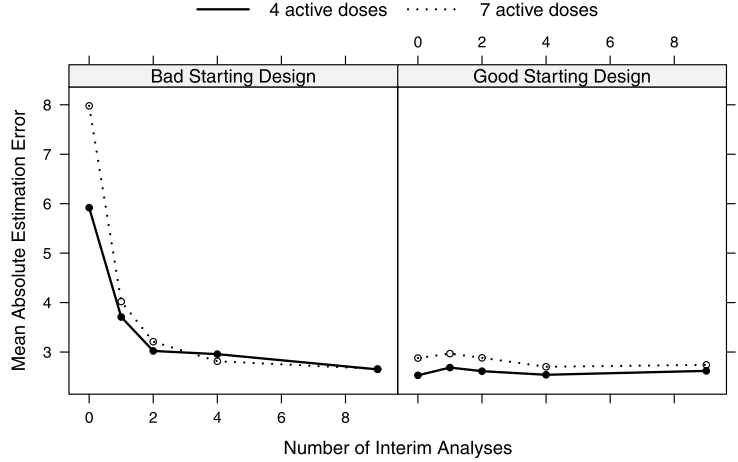}
\vspace*{-5pt}
\caption{Mean absolute estimation error for MED estimation,
under the the Logistic$_1$ model for good and bad starting designs.}
\label{fig4}
\vspace*{-3pt}
\end{figure}

The adaptive design benefits observed so far depend on several input
parameters, such as the starting design for the first
stage. One may argue that starting with a bad design that allocates
patients at the ``wrong'' doses may be improved by adapting at one or
more interim looks. On the other hand, starting with a good design may
lead to adaptations following random noise at the interim analyses. To
illustrate this effect, we report the results for the simulations under
the Logistic$_1$ model (similar results were also obtained for other
models and scenarios, but are not reported here). We used four
different starting designs. We used ${\bm w} = (0.35, 0.03, 0.22,
0.35, 0.05)$ and ${\bm w}= (0.35, 0.02, 0.02, 0.02, 0.02, 0.20, 0.30, 0.07)$
as good starting designs with 4 and 7 active doses,
respectively. These designs work well because they allocate patients
on placebo and around the MED, while keeping some mass on the
remaining doses. In addition, we used ${\bm w} = (0.1, 0.3, 0.05,
0.05, 0.5)$ and ${\bm w} = (0.1, 0.2, 0.22, 0.02, 0.02, 0.02, 0.02,
0.4)'$ as bad starting designs, as they have relatively few patients on
placebo and
around the MED. It follows from Figure~\ref{fig4} that substantial
improvements are possible when using bad starting designs. On the
other hand, for good starting designs no benefit is achieved by
adapting and the performance may even deteriorate, because the
possibility of adapting may lead one to deviate from the already good
starting design. In practice, one does not know whether an
employed design is good or bad, but one should keep in mind the
possibility that adaptive designs will not always improve
upon the initial design.

To further investigate the robustness of the proposed methods, we
repeated the simulation study from Figure \ref{fig2} by
increasing the standard deviation to 450 ml and 700 ml. The overall results
remain similar, but with increased absolute
estimation errors. However, the relative benefit of 9 interim analyses
vs. no adaptation decreases slightly. Due to the larger noise, one
obtains less reliable information at an interim analysis and one may
end up with a worse design for the next stage. We also investigated
the effect of prior misspecification. For this purpose we misspecified
the prior means or prior modes by adding or subtracting 20\% of the
true value, but leaving the variability (variance of baseline and
maximum effect and the value $S$ for the beta distribution) unchanged
as in the original simulations. The results are largely identical to
those presented in Figure \ref{fig2}, indicating that the proposed
methods are robust under moderate prior misspecifications.

\subsection{Conclusions for asthma study}

Many more simulations than presented above were conducted at the
planning stage of the asthma study to address the four questions
stated in Section~\ref{sec:example}. Regarding question (A), it was
felt that the potential benefits of conducting an adaptive design
(more precise MED estimation) outweighed the additional logistical
requirements, especially in view of the perceived sample size gain of
100--200 patients when compared to a fixed-sample study designed to
achieve a similar precision. For question (B) it was decided to have
one interim analysis: based on Figure~\ref{fig2} and other simulation
results, the potential further reduction of the mean absolute
estimation error with two or more interim analyses was perceived as
too small to justify the additional logistical complexity.

For similar reasons, it was decided against having all seven actives
doses from the beginning on question (C). Instead, 150 patients
ought to be allocated equally across the four active doses 2.5, 10,
20, 50 (plus placebo) in the first stage. Once the interim results
are available and analyzed with the methods from
Section~\ref{sec:meth}, however, patients could be allocated to all
seven active doses (or a subset thereof) in the second stage. For
practical reasons, the clinical team decided to incorporate
constraints on the minimum number of patients allocated per dose in
the second stage: if the algorithm would allocate less than 5\% of the
patients on a certain dose, that dose would be dropped altogether and
the corresponding patients reallocated to the remaining doses.

\section{Discussion}

Motivated by a dose finding study in moderate persistent asthma, we
described a response-adaptive approach that addresses common
challenges encountered in dose-finding studies: multiple objectives,
model uncertainty, and large variability. When planning an adaptive
dose-finding design it is important to realize that it may not always
be better than a nonadaptive design. It is necessary to employ a
factored view, as many parameters may impact the performance of a study
design. Often, an unbalanced fixed-sample design derived from optimal
design theory might already provide benefit over a balanced
fixed-sample design and adaptation may not bring further advantages,
particularly if the variability is large (which is common in
practice). Thus, adaptive designs are promising in situations where
the initial design is not good and/or interim parameter estimates have low
variability. In practice, one never knows how good the initial design
will be, before trial start, and adaptive designs may guard against
bad initial designs. However, the benefits of adaptive dose-finding
designs have to be balanced against the increased logistical
requirements to implement processes for repeated data collection,
cleaning and analyses, to maintain trial integrity and validity, and
to overcome potential challenges in drug manufacturing and supply.

In this paper we focused on designs based on the compound optimality
criterion \eqref{eqn:optCrit2} to address model uncertainty and to
minimize the variance of $\widehat{\mathrm{MED}}$. The criterion
depends on the parameters of the different dose-response models as
well as on the model probabilities and we used a Bayesian approach to
continuously update parameter values and model probabilities based on
the information accrued in the trial. The approach was implemented
based on optimization and numerical quadrature, so that
computationally intensive Markov chain Monte Carlo techniques could be
avoided. Computational efficiency is of extreme importance, as the
frequentist operating characteristics of any adaptive design
methodology needs to be evaluated in extensive simulations under
multiple scenarios.

The proposed method can be extended immediately if alternative
optimality criteria [such as ED$_p$- or D-optimal designs, see
\citet{dett2010}] or mixtures thereof are of interest.
Alternatively, optimal discrimination designs could be applied that
allow one to differentiate among several candidate nonlinear
regression models [\citet{atkifedo1975},
\citet{detttito2009}]. It would be interesting to address
multiple objectives by considering different optimality criteria at
different stages, such as using a model discrimination design in
earlier stages, and MED-optimal design in later stages. This will be
investigated in future research, but see \citet{drag2010} for
initial results.

The R functions used for the simulations are available with the\break
\texttt{DoseFinding}~R package [see \citet{bornpinhbret2010}].

\begin{appendix}

\section*{\texorpdfstring{Appendix: Proof of Theorem~\lowercase{\protect\ref{theo1}}}{Appendix: Proof of Theorem 3.1 }}\label{app}\label{appm}

Obviously the first part of the theorem
follows from the lower bound \eqref{low} on the efficiency. For a
proof of \eqref{low} let $\gamma= N_{\mathrm{old}}/(N_{\mathrm{old}}+N_{\mathrm{next}}) \in
(0,1)$ and note that the total information of the experiment in the
$m$th model is given by
%
\begin{equation}
\label{a1}
\hspace*{20pt}\bm M_m (\bolds\theta, \bm w_{\mathrm{old}}, \bm
w_{\mathrm{next}})= \gamma\bm M_m (\bolds\theta_m, \bm w_{\mathrm{old}})+(1-\gamma)\bm
M_m (\bolds\theta_m, \bm w_{\mathrm{next}}),
\end{equation}
where we collect in the vector $\bolds{\theta}=(\bolds\theta_1,\ldots ,\bolds
\theta_M)$ the parameters of the different models. Define a block
diagonal matrix by
%
\begin{eqnarray}\label{a2}
&&\overline\bm{ M} (\bolds\theta_m, \bm w_{\mathrm{old}}, \bm w_{\mathrm{next}})\nonumber
\\[-8pt]
\\[-8pt] && \qquad = \operatorname
{diag} (\bm{M}_1(\bolds\theta_m, \bm w_{\mathrm{old}}, \bm w_{\mathrm{next}}),\ldots ,
\bm{M}_M(\bolds\theta_M, \bm w_{\mathrm{old}}, \bm w_{\mathrm{next}}))
\nonumber
\end{eqnarray}
(all other entries in this matrix are 0) and, similarly,
\[
K= \operatorname{diag} (c_1(\bolds\theta_1),\ldots ,c_M(\bolds\theta_M)),
\]
where the vector $c_m(\bolds\theta)$ is given by $c_m(\bolds\theta) =
\bolds
\nabla b_m (\bolds\theta), m=1,\ldots ,M$. For a design $\bm w_{\mathrm{next}}$,
such that $c_m(\bolds\theta) \in\operatorname{Range} (\bm{M}_m(\bolds\theta_m,
\bm w_{\mathrm{old}},\bm w_{\mathrm{next}}))$ $(m=1,\ldots ,M)$, we consider the information
matrix
\begin{eqnarray*}
C_K (\overline\bm{ M} (\bolds\theta, \bm w_{\mathrm{old}}, \bm w_{\mathrm{next}})) &=&
(K^T \overline\bm{ M}^{\,-} (\bolds\theta, \bm w_{\mathrm{old}},\bm
w_{\mathrm{next}})K)^{-1} \\
&=& \operatorname{diag} ((c^T_1 (\bolds\theta_1) \bm M^-_1 (\bolds\theta_1, \bm
w_{\mathrm{old}}, \bm w_{\mathrm{next}}) c_1(\bolds\theta_1))^{-1}, \ldots , \\
&&\hspace*{28pt} (c^T_M(\bolds\theta_M)\bm M^-_M (\bolds\theta_M, \bm w_{\mathrm{old}}, \bm
w_{\mathrm{next}}) c_M(\bolds\theta_M))^{-1}).
\end{eqnarray*}
Note that the optimal design maximizes
\begin{eqnarray*}
\Psi^{-1} (\bm w_{\mathrm{next}}) &=& \frac{N_{\mathrm{old}} + N_{\mathrm{next}}}{\sigma^2}
\cdot\Phi_\alpha
(C_K (\overline\bm{ M}(\bolds\theta, \bm w_{\mathrm{old}}, \bm w_{\mathrm{next}}))) \\
&=& \frac{N_{\mathrm{old}}+N_{\mathrm{next}}}{\sigma^2} \prod^M_{m=1} (c^T_m (\bolds
\theta) \bm M^-_M (\bolds\theta_m, \bm w_{\mathrm{old}},
\bm w_{\mathrm{next}})c_m(\bolds\theta))^{- \alpha_m},
\end{eqnarray*}
where the last identity defines the criterion $\Phi_\alpha$ and we
have used the notation
$\Phi_\alpha(\operatorname{diag}(\lambda_1, \ldots , \lambda_M))=\prod
^M_{m=1} \lambda^{\alpha_m}_m$.
Now according to Theorem~1 in \citet{dett1996}, a lower bound for the
efficiency of the design $\bm w_{\mathrm{next}}$
\[
r(\bm w)=\frac{\psi^{-1}(\bm w)}{\psi^{-1}(\bm w_{\mathrm{opt}})}=
\frac{\Phi_\alpha(C_k (\overline\bm{ M} (\bolds\theta,\bm w_{\mathrm{old}},
\bm w_{\mathrm{next}}))) }
{\max_{\bm v \in\mathbb{S}^k} \Phi_\alpha(C_K (\overline\bm{ M}
(\bolds\theta, \bm w_{\mathrm{old}}, \bm v)))}
\]
is obtained as
\begin{eqnarray*}
&&e=  \Bigl[ \min_{\bm G } \max_{A \in\mathcal{A}}   \operatorname{tr}
\{ \bm
GKC_K(\overline\bm{ M} (\bolds\theta, \bm w_{\mathrm{old}}, \bm w_{\mathrm{next}}))\\
&&\hphantom{e=  \Bigl[ \min_{\bm G } \max_{A \in\mathcal{A}}   \operatorname{tr}
\{}
{}\times\bm
EC_K (\overline\bm{ M} (\bolds\theta, \bm w_{\mathrm{old}}, \bm w_{\mathrm{next}})) K^T
\bm G^T
A  \}  \Bigr]^{-1},
\end{eqnarray*}
where the minimum is taken over the set of all generalized inverses of the
matrix $\overline\bm{ M}(\bolds\theta, \bm w_{\mathrm{old}}, \bm w_{\mathrm{next}})$ and
the set $\mathcal{A}$ is defined by
\[
\mathcal{A} = \{ \overline\bm{ M } (\bolds\theta, \bm w_{\mathrm{old}}, \bm v)
\mid\bm v \in\mathbb{S}^k \}
\]
and the matrix $\bm E$ is given by
\[
\bm E= \operatorname{diag} (\alpha_1 c^T_1 (\bolds\theta) \bm M^-_1(\bolds\theta
_1, \bm w_{\mathrm{old}}, \bm v)c_1(\bolds\theta),\ldots , \alpha_M c^T_M (\bolds
\theta)
\bm M^-_M (\bolds\theta_M, \bm w_{\mathrm{old}}, \bm v) c_m).
\]
Therefore, observing the identity
\[
\bm M_m(\bolds\theta_m, \bm w_{\mathrm{old}}, \bm v) = \bm M_m(\bolds\theta_m,
\bm w_{\mathrm{old}}, \bm w_{\mathrm{next}})+ (1-\gamma)
\bigl(\bm M(\bolds\theta_m,\bm v)- \bm M(\bolds\theta, \bm w_{\mathrm{next}})\bigr),
\]
we obtain
\begin{eqnarray*}
e&=&  \Biggl[ 1+(1-\gamma) \min_{\bm G_m } \max_{v \in\mathbb{S}^k}
\sum^M_{m=1} \alpha_m
c^T_m(\bolds\theta_m)\bm G^T_m \bigl(\bm M_m(\bolds\theta_m,\bm v)-\bm
M_m (\bolds\theta_m, \bm w_{\mathrm{next}})\bigr)\\
&&\hphantom{\biggl[ 1+(1-\gamma) \min_{\bm G_m } \max_{v \in\mathbb{S}^k}
\sum^M_{m=1}}\hspace*{45pt}
{}\times\bm G_m c_m(\bolds\theta)
/
(c^T_m (\bolds
\theta_m) \bm G_m c_m (\bolds\theta_m))  \Biggr]^{-1}
\\
&\geq&  \Biggl[ \min_{\bm G_1,\ldots ,\bm G_m} \max_{d \in\{d_1,\ldots
,d_k \}}
\sum^M_{m=1} \alpha_m \frac{(g^T_m(d, \bolds\theta
_m)\bm G_m
c_m (\bolds\theta_m))^2} {c^T_m (\bolds\theta_m)\bm G_m c_m (\bolds\theta
_m)}\\
&&\hphantom{\biggl[ \min_{\bm G_1,\ldots ,\bm G_m} \max_{d \in\{d_1,\ldots
,d_k \}}}
{}\Big/
\sum^M_{m=1} \alpha_m \frac{c^T_m (\bolds\theta_m)\bm G^T_m M_m
(\bolds
\theta_m, \bm w_{\mathrm{next}}) \bm G_m c_m (\bolds\theta_m)} {c^T_m(\bolds\theta
_m) \bm G_m c_m (\bolds\theta_m)}
  \Biggr]^{-1},
\end{eqnarray*}
where we have used the inequality
\[
[1+(1-\gamma)(A-B)]^{-1} \geq\biggl[\frac{A}{B}\biggr]^{-1}
\]
for $A \geq B \ge0 $, $(1-\gamma) B \leq1$ and standard arguments in
design theory.
\end{appendix}

\section*{Acknowledgments}
The authors would like to thank Martina Stein,
who typed parts of this manuscript with considerable technical
expertise.


%
\vspace*{-0.8pt}

\printaddresses

\end{document}